\documentclass[12pt,preprint]{aastex}
\begin{document}

\newcommand{\pc}{{\rm pc}}
\newcommand{\msun}{{M_\odot}}
\newcommand{\Msun}{{M_\odot}}
\newcommand{\Lsun}{{L_\odot}}
\newcommand{\yr}{{\rm yr}}
\newcommand{\kms}{{\rm km}{\rm s}^{-1}}
\newcommand{\beq}{\begin{equation}}
\newcommand{\beqa}{\begin{eqnarray}}
\newcommand{\eeq}{\end{equation}}
\newcommand{\eeqa}{\end{eqnarray}}
\title{Galaxy Bulges As Tests of CDM vs MOND in Strong Gravity}
\author{HongSheng Zhao\altaffilmark{1,2}, Bing-Xiao Xu\altaffilmark{3,4},  Clare Dobbs\altaffilmark{5} }
\email{hz4@st-andrews.ac.uk}
\altaffiltext{1}{SUPA, University of St Andrews, KY16 9SS, Fife, UK}
\altaffiltext{2}{National Astronomical Observatories, Chinese Academy of Sciences, 20A Datun Road, Chaoyang District, Beijing, China}
\altaffiltext{3}{Department of Astronomy, Peking University, 100871,Beijing, China}
\altaffiltext{4}{Department of Physics and Astronomy, Georgia State University, Atlanta, GA
30303, USA}
\altaffiltext{5}{School of Physics, University of Exeter, UK}


\begin{abstract}
The tight correlation between galaxy bulges and their central black hole masses likely emerges in a phase of rapid collapse and starburst at high redshift, due to the balance of gravity on gas with the feedback force from starbursts and the wind from the black hole; the average gravity on per unit mass of gas is $\sim 2 \times 10^{-10}{\rm m}\ {\rm sec}^{-2}$ during the star burst  phase.   This level of gravity 
could come from the real $r^{-1}$ cusps of Cold Dark Matter (CDM) halos, but the predicted gravity would have a large scatter due to dependence on cosmological parameters and formation histories.
Better agreement is found with the gravity from the scalar field in some co-variant versions of MOND, which can create the mirage of a Newtonian effective dark halo of density $\Pi r^{-1}$ near the center, where the characteristic surface density $\Pi=130\alpha^{-1} \msun\pc^{-2}$  and $\alpha$ is a fundamental constant of order unity fixed by the Lagrangian of the co-variant theory if neglecting environmental effects.  

We show with a toy analytical model and a hydrodynamical simulation that a constant background gravity due to MOND/TeVeS scalar field implies a critical pressure synchronizing starbursts and the formation of galaxy bulges and ellipticals.  A universal threshold for the formation of the brightest regions of galaxies in a MONDian universe suggests that the central BHs, bulges and ellipticals would respect tight correlations like the $M_{bulge}-M_{BH}-\sigma$ relations. In general MOND tends to produce tight correlations in galaxy properties because its effective halo has less freedom and scatter than CDM halos.  
\end{abstract}

\maketitle

\section{Correlated formation of Black Hole and Bulges in CDM and MOND}

While appearing in wide range of shapes, sizes and luminosities, galaxies have very regular properties.   E.g., the terminal rotation speed $V_{cir}$ of a spiral galaxy is tightly correlated with its total baryonic mass $M$, following a simple TFMM power-law $V_{cir}^4/M \sim 0.02 (\kms)^{-4}\Msun^{-1}$, a formula proposed by Tully \& Fisher (1977) for high-surface brightness galaxies, and generalized by Milgrom (1983) and tested by McGaugh (2005) for gas-rich low-surface brightness galaxies.  There is no evidence for any significant scatter, and it seems to apply independent of galaxy formation history (Gentile et al. 2007).  This power-law also applies, with significant scatter, to elliptical galaxies and bulges if replacing $V_ {cir}$ with $ \sim 1-2$ times the typical stellar dispersion $\sigma$ (Faber \& Jackson 1976).  
Nevertheless, a much tighter relation exists for the central black holes of these nearly spherical systems.

The formation of central black holes (BHs) in galaxies is likely to be a
rapid process since most quasars have already formed at redshift
$z>2$. There is a tight correlation between the BH mass and 
the mass of the spheroidal (bulge) component, or even better 
its velocity dispersion  (Ferrarese \&
Merritt 2000; Gebhardt et al. 2000; Tremaine et al. 2002).   
The correlations $M_{bulge} \propto M_{BH} \propto \sigma^4$ are so 
tight that it is hard to explain unless bulges form as fast as BHs,
and their growth is controlled simultaneously by some mechanism.
At the present epoch, the BH accretion rate is both small and completely decoupled from the bulge growth. The likely window to couple the two is at high redshift   
during phases of rapid growth and violent feedback.  
Many previous discussions emphasize that the feedback
from central supermassive black hole, which interacts with the
surrounding environment in a self-regulated way, is the key to form
the correlations (Silk \& Rees 1998; King 2003; Wyithe \& Loeb
2003; Murray, Quataert \& Thompson 2005; Begelman \& Nath 2005;
Cen 2007). On the other hand, the starburst activities peak at
similar redshifts to the quasars as a whole. In a co-evolution
scenario of starbursts and a central BH, the central BH accretes
with high accretion rate during the main star formation (SF) phase
(Alexander 2005).  To make the starbursts, it was proposed that
bulges can form by a rapid collapse due to radial instability of
isothermal gas. This proposal has the nice feature of forming bulges
before disks (Xu, Wu, \& Zhao 2007).  Inspired by these
works in a Newtonian framework, we model the criteria of bulge formation, assuming a more general mixture of gas and a non-isotropic stellar component imbeded in a constant external gravity provided by either CDM halos or effective halos of a MOND scalar field.

Observations show that most of the local and distant starburst galaxies are rich in gas and dust
(Heckman, Armus, \& Miley 1990; Meurer et al. 1995; Sanders
\& Mirabel 1996; Adelberger \& Steidel 2000). 
Photons from newly-formed stars and the BH's accretion disk 
with a luminosity $L_{SF}$ and $L_{BH}$ would diffuse out of the gas sphere.  While keeping the gas nearly isothermal, the photons exert a pressure due to dust opacity.  
The momentum deposit rate of photons from an Eddington accreting BH, 
$(2/c)L_{BH} = (M_{BH}/\msun) \times 4.3 \times 10^{-8} {\rm m s}^{-2} \msun  = (M_{BH}/1.2 \times 10^8\msun) \times 10^{31}$ Newton, might drive a feedback force on the gas (King \& Pounds 2003) to balance the inward momentum deposit of the SF, 
$(2/c)L_{SF}=(L_{SF}/3.7 \times 10^{12}L_\sun) \times 10^{31}$ Newton; the latter could also drive an outward force to balance, say half of, the gravitational force on the gas.  So we have roughly 
\beq 
(10^{-10} {\rm m s}^{-2}) (430 M_{BH})  \sim {2\over c} L_{BH} \sim {2 \over c} L_{SF} \sim 0.5\bar{g} M_{g} 
\eeq
where $\bar{g}$ is the mass-averaged gravity on the gas sphere.  Rewriting $\bar{g}=10^{-10}{\rm m s}^{-2} g_{-10}$, we find a short star formation time scale 
$(0.001 c^2)M_g/L_{SF} \sim 0.2 g_{-10}^{-1}$Gyr 
if adopting the usual SF efficiency of about 0.001 (Leitherer et al. 1999; Bruzual
\& Charlot 2003).  In a starburst, we expect radial collapse and violent feedback, so 
the short timescale of SF should be comparable to the free-fall time scale. 
Assuming that all the gas eventually turned into stars, we recover a Magorrian et al.  (1998) relation ${M_{BH} \over M_*} \sim 0.002 g_{-10} $.  The key point here is that observed properties of BH, SF and bulges can all be realised {\it if $g_{-10} \sim 2$ with little scatter universally}.

Presently there are two paradigms where galaxy structure and formation are studied.  Introducing only two speculative dark components of the universe, the paradigm of Cold Dark Matter (CDM) plus the cosmological constant $\Lambda \sim (10^{-9}m/s^2)^2$, it is possible to simulate the large scale universe realistically with General Relativity. Many properties of model galaxies can be predicted, although not all are in agreement with observations, especially for low-surface brightness galaxies.  The Modified Newtonian Dynamics (MOND) paradigm is able to match properties of high and low surface brightness galaxies with amazing accuracy by simply introducing a fundamental scale $a_0 \sim 1.2 \times 10^{-10}m/s^2$ in the space-time metric gradient.  However, it is generally underdeveloped in terms of simulating the processes of structure and galaxy formation.  Nevertheless, it was realised that disks in MOND become unstable once above certain central surface brightness ${a_0 \over 2\pi G} \sim 130 \msun\pc^{-2}$, so bright regions of galaxies tend to exist in spheroidal form and are in strong gravity $g\ge a_0$ (Sanders \& McGaugh 2002); throughout the paper we use $a_0$ as the dividing scale for strong vs. weak gravity.     

The classical MOND gravitational theory (Bekenstein \& Milgrom 1984)
is now furbished with several co-variant versions.  These have different constructions using a vector field plus (optional) scalar fields (Bekenstein 2004, Sanders 2005, Halle, Zhao, \& Li 2008, Zhao \& Li 2008); the scalar field(s) can always be absorbed (e.g., into the modulus of the vector field), but can be useful to make the expressions of the theories simpler.  
In these theories there is a constant scale $a_0 \sim 1.2 \times 10^{-10}{\rm m}\ {\rm sec}^{-2}$; in regions or epochs of more gradual variation of the space-time metric,
the dominating source to Einstein's tensor 
switches from normal matter/radiation fields to the new vector or scalar fields.
With the same scale the most recent co-variant model (Zhao 2007) can explain how 
the expansion of the universe switches from de-accelerating to accelerating, 
and explains the amplitude of the cosmological constant $\Lambda \sim (8 a_0)^2$, rather than invoking it arbitrarily as in $\Lambda$CDM.  
In principle, cosmological structure formation all the way to galaxies is
well-defined in some of these co-variant theories (Halle, Zhao \& Li 2008).  In the original
proposal for TeVeS (Bekenstein 2004), there is a dis-continuity in the Lagrangian for the scalar field, which makes the evolution from cosmological linear perturbations to quasi-static galaxies problematic.  This is overcome in the proposed Lagrangian of 
Zhao \& Famaey (2006).  It is theoretically possible to seed and simulate galaxy formation from the linear perturbation simulations of the Cosmic Microwave Background (Skordis et al. 2005).  Already work has started on galaxy formation modeling in MOND (Sanders 2008, Sanders \& Land 2008) and there are many constraints on the theory from gravitational lensing tests (Feix et al. 2007, 2008, Shan, Feix, Famaey, \& Zhao 2008, Natarajan \& Zhao 2008, Tian, Hoekstra, Zhao 2008). 

Here we speculate upon the properties of galaxies formed in the TeVeS framework, and contrast with the $\Lambda$CDM framework.
While generally speaking the two paradigms are mutually exclusive, here we illustrate some similarities, in the context of the formation of high surface brightness bulges and elliptical galaxies.  There are, however, important differences.

\section{Scatter of the halo gravity in CDM}

\begin{figure}
\includegraphics[height=16cm, width=16cm,angle=0]{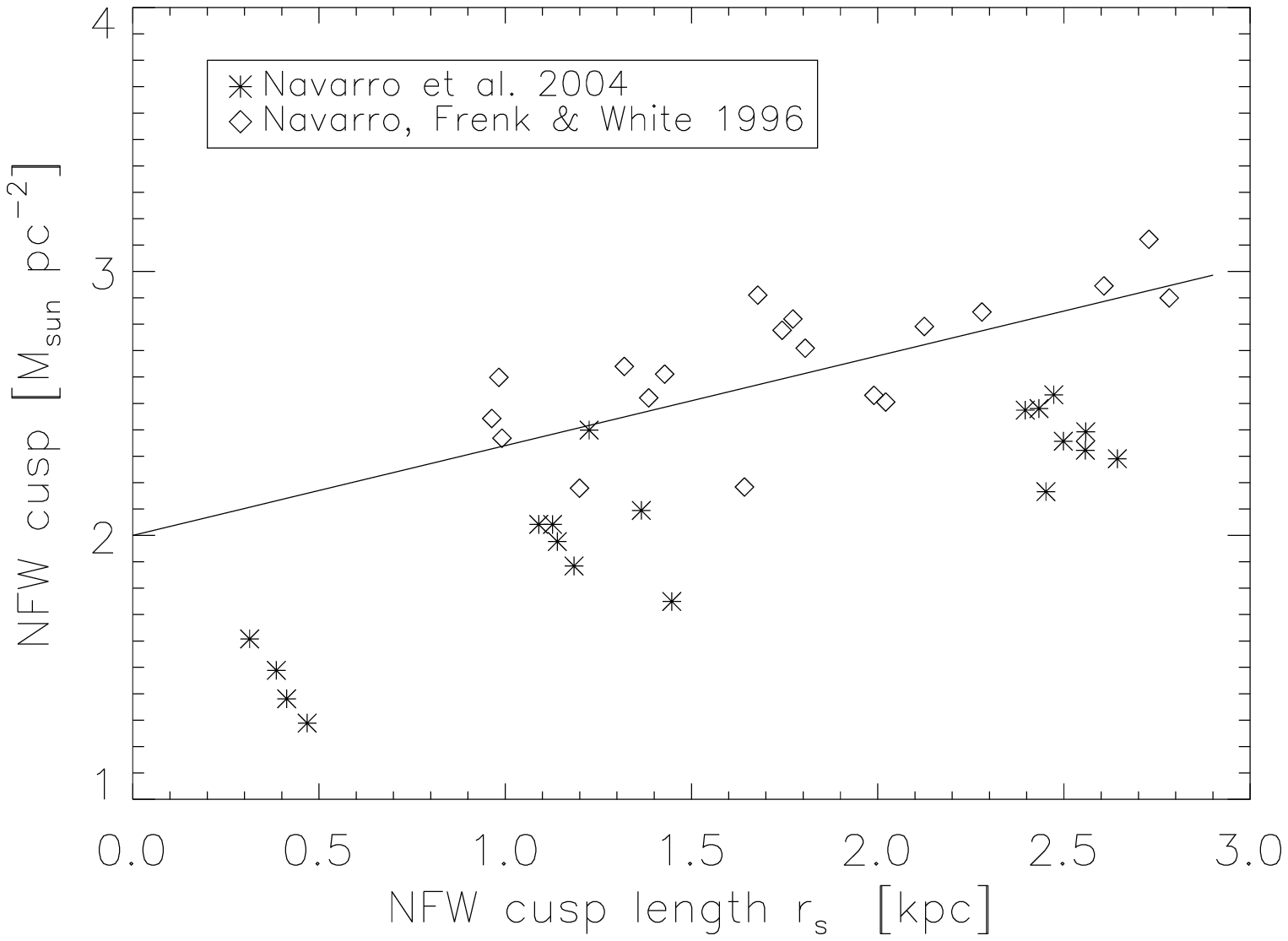}
\vskip 3cm
\caption{Shows the scatter in $\log r_s$ in ${\rm kpc}$ vs 
 $\log \rho_s r_s$ in $\msun\pc^{-2}$ in simulated NFW halos.  The line shows 
our empirical relation $\rho_s r_s = 130\Xi \msun\pc^{-2}$, and $\Xi=2(r_s/17h^{-1}{\rm kpc})^{0.34}$ (cf. eq.5).  The data points from Navarro et al. (2004, adopting $\Omega_\Lambda= 0.7$ and $\sigma_8=0.9$, $h=0.7$)  are systematically below those from the original NFW (1996, adopting $\Omega_\Lambda=0$, $\sigma_8=0.63$ and $h=0.5$); apart from scatter, the halo concentration has systematic dependence both on the cosmology and on the halo size.  In comparison MOND predicts a universal effective DM scale $\log (r \rho) \sim \log 130 \sim 2.3$ (not shown). }\label{scatter}
\end{figure}

Let us consider the Cold Dark Matter (CDM) galaxy formation framework.
In CDM models baryons fall into the potential well of CDM, cool, and condense into stars.  The background dark matter distribution is often described
by the NFW density distribution for dark matter (Navarro, Frenk \& White 1997, Navarro et al. 2004 ), which decreases smoothly from an $r^{-1}$ density cusp inside to a $r^{-3}$ tail outside a scale radius $r_s$,
defined to be where the logarithmic density slope is $-2$ exactly. The density satisfies $ r \rho_{NFW} \approx \rho_s r_s$, insensitive to radius inside the cusp region $r<r_s$, where $\rho_s$ is a density scale.
The Newtonian gravity of the halo, determined by the halo mass enclosed inside radius $r$, is given by
\beq
g_{NFW}(r) = {G M_{NFW}(r) \over r^2} = {(2 \pi G) (\rho_s r_s)} F\left({r \over r_s}\right),
\eeq
where the function
\beq
F(y)  \equiv {2 \over y^2} \left[\ln(1+y)-{y \over 1+y}\right] \sim (1+y)^{-1.475}.
\eeq
The approximation is valid to 10 percent for $0<y<20$ and gives $F(0)=1$.    For a halo of virial mass $M_{vir}$, the halo scale parameters satisfy
\begin{eqnarray}
r_s &=& (17 {\rm kpc/h}) M_{vir,12}^{0.46}, \\
r_s \rho_s &=& 130 \msun\pc^{-2} \times \Xi, \\
\Xi & =&  2 M_{vir,12}^{0.17} \sim 2 \left({r_s \over 17/h {\rm kpc}}\right)^{0.34}.
\end{eqnarray}
where $h \equiv H_0/100$ is the dimensionless Hubble constant, and 
the halo virial mass $M_{vir}$ is related to the baryonic mass $M_b$ of a galaxy 
by $M_{vir} \sim 8 M_b$.   
The parameters are insensitive to the resolution of the simulation
and whether the cusp is truly finite or infinite at the origin.
The numerical values above and the data points shown in Fig.~\ref{scatter} 
are taken from Navarro et al. (1996) and
Table 3 of the simulations of Navarro et al. (2004).
They agree with the scalings for
the concentration vs mass, $c  \equiv {r_{vir} \over r_s} \sim  13.4 M_{vir,12}^{-0.13} (1+z)^{-1}$, as given in Bullock et al. (2001), where $c$ is a ratio betwen the virial radius $r_{vir}$ and the scale radius $r_s$.

It is interesting that in the cusp, $F \sim 1$, and the NFW halo's self-gravity $g_{NFW}(r)$ is insensitive to radius and halo parameters, i.e.,
\beq
g_{NFW}(r)  \sim (1.2 \pm 0.6) \Xi \times 10^{-10} m/s^2, 
\eeq
hence $g_{NFW}(r)$ of NFW cusps is nearly a universal constant for galaxies of the same baryonic total mass.  A factor of two scatter in $g_{NFW}(r)$ is estimated from   
Table 3 and Fig.1a of Navarro et al. (2004); the galaxy clusters of $10^{14}\msun$ have nearly
the same $r^{-1}$ extrapolated inner density as $10^{12}\msun$ galaxy halos (although the cluster simulations do not yet have enough resolution to be confidently extrapolated to small radii of a few kpc). 

More precisely $r_s^{0.66} \rho_s \sim cst$ insensitive to halo mass, and simulation resolution, i.e., it applies
to the $r^{-1.5}$ Moore profile as well the cored profile of Navarro et al. (2004) as long as $r_s$ has the model-insensitive definition of the radius where the logarithmic density slope is $-2$.  The cluster density is a factor of two higher, and the dwarfs are a factor of four lower due to the $\Xi$ factor.  

The above is roughly in agreement with Xu et al. (2007), who noted that
in the central region containing the galaxy bulge, NFW halos have a density scaling relation $r \rho \sim r_s \rho_s \sim 130 \msun\pc^{-2} \Xi$, and $\Xi(M_{vir},z,c) \sim M_{vir,12}^{0.072} \sim 1$.  While the details differ, in both cases $\Xi$
is a shallow function of the halo virial mass $M_{vir}$, the redshift $z$, and the concentration $c$.

\subsection{Effects on CDM cusp if bulges and ellipticals grow adiabatically}
The CDM density is highly compressible by gravitational interaction with baryons.  The stellar distribution in elliptical galaxies is often described by Sersic profile in projected light (see Graham \& Driver 2005), with an underlying volume density of the approximate form $r^{-1+0.6/n} \exp(-r^{1/n})$ (Prugniel \& Simien 1997); for the Sersic index $n \sim 4$, we have a nearly $1/r$ cusp, which suggests a nearly constant Newtonian gravity in the center.  Here we use the simpler Hernquist profile for the enclosed mass $M_b(r)$, hence we find the central Newtonian gravity has a spatially constant value,
\beq\label{Hernquist}
g_N(r)={G M_b(r) \over r^2}= {G M_b \over (r+r_h)^2} \sim {G M_b \over r_h^2},
\eeq 
for $r \rightarrow 0$, where $M_b$ and $r_h$ are the baryon (total) mass and scale length.  
The scale length and stellar mass (luminosity) of elliptical galaxies are typically correlated with some scatter, e.g., Chen \& Zhao (2008) find that 
\beq
\log {G M_b r_h^{-2} \over 350 \times 10^{-10}ms^{-2} } = -1.52 \log \left({M_b \over 1.5 \times 10^{11}\msun}\right) \pm 0.5.  
\eeq
where $1.5 \times 10^{11}M_{\sun}$ is the characteristic turn-over mass scale in the Schechter stellar mass function of galaxies, found by fitting SDSS galaxy samples in the range $10^8-10^{12}\msun$ (Panters et al. 2004), assuming a Hubble parameter $70$km/s/Mpc.  
This would imply that
near the Hernquist cusp the Newtonian gravity in units of $10^{-10}m/s^2$ is
$g_{N-10} = 350 \times \left({M_b \over 1.5 \times 10^{11}\msun}\right)^{-1.52}$.  This illustrate that cores of ellipticals have $g \gg a_0$, i.e., they are strong gravity enviornment (Milgrom \& Sanders 2003). 
\footnote{As an alternative model-insensitive check, we estimate from Faber et al. (1997, their Eq.3-4, and Fig.4) that the typical observed Newtonian gravity near the core or the half-mass radii of cored giant ellipticals (with $L_{10} \equiv L/10^{10}\Lsun > 1$) is
$g_{N-10} \sim 100 L_{10}^{-0.6}$, and for cusped dwarf ellipticals (with $L_{10}<1$)
is $g_{N-10} \sim 1000 L_{10}^{+0.6}$.} 

{\it If} in elliptical galaxies baryons condense into the center adiabatically, this process would further increase the value of $r \rho_{NFW}$ or $\Xi$.  In the most extreme case, a galaxy might start with 
a $f_b:(1-f_b)=1:7$ mix of gas and CDM particles all distributed on circular orbits with an NFW distribution of the radius $r_i$, and the gas part condense adiabatically into the present stellar mass distribution $M_b(r)$.  Following the standard recipe (e.g., Klypin, Zhao, Somerville 2002), the initial CDM halo mass $(1-f_b)M_{NFW}(r_i)$ is squeezed into a radius of $r$, conserving mass.  From conservation of angular momentum $J$ of circular orbits, we have 
\beq
J^2= r_i GM_{NFW}(r_i) = r \left[(1-f_b)  GM_{NFW}(r_i) + GM_b(r) \right].
\eeq
By Taylor expanding near $r \sim 0$, we find the contraction factor $C \equiv {r_i \over r}$ satisfies
\beq
C^3 \approx (1-f_b) C^2 + {G M_b r_h^{-2} \over g_{NFW}(0)  } 
 \sim 350 \left({M_b \over 1.5 \times 10^{11}\msun}\right)^{-1.52} \!\!\!\!\!\! \gg 1,
\eeq
where we considered only the dominant 2nd term and applied the approximation of Chen \& Zhao (2006).
By contracting the same halo mass inside a factor of $C$ smaller radii, one keeps the $r^{-1}$ profile of of the halo cusp, but 
increases the halo central density and self-gravity by a factor $C^2$, i.e., at the center 
\beq
g_{CDM} = 2 \pi G r \rho_{CDM} = C^2 g_{NFW} = C^2 \Xi \cdot 10^{-10}m/s^2 \gg a_0.
\eeq
The above is likely an over-estimate, since more realistic formation of ellipticals will involve mergers of gas-rich spirals galaxies.  In any case, there is likely a {\it large upward scatter} around the naive analytical prediction $g_{CDM} \sim 10^{-10}m/s^2$ due to different scenarios of formation and different halo masses and concentrations etc..

\section{A universal gravity scale: effective halos in simple MOND}

In the co-variant theories of MOND, the only sources
of gravity are the stars and the gas (i.e., the baryons). In the context of
its co-variant version, the Tensor-Vector-Scalar theory of Bekenstein
(2004), there should be a scalar field $\phi_s$, such that in the
spherical case ${\mathbf g}_s = -\nabla \phi_s$ gives a dark halo like
gravity.  Here we call the scalar field the Effective Dark Matter (EDM),
$g_{EDM}=g_s=|{\mathbf g}_s|$; it is
related to the Newtonian gravity $g_N=|{\mathbf g}_N|$  and the actual acceleration (or gravity) $g=|{\mathbf g}|=|\mathbf{g}_N+\mathbf{g}_s|$.  The Poisson equation is modified as
\beq
 \nabla \cdot (\mu_s \mathbf{g}_s ) = \nabla \cdot (\mu {\mathbf g}) = \nabla \cdot {\mathbf g}_N = -4 \pi G \rho,
\eeq
where $\mu_s$ and $\mu$ are modification functions, which give
identical descriptions in case of a spherical mass distribution, and reduce to Newtonian dynamics when $g_N \gg a_0$.

Consider the modification functions as proposed in Angus et al. (2006),
\begin{equation}
\mu_s = {g_s \over (a_\alpha - g_s)\alpha},\qquad 
a_\alpha  \equiv {a_0 \over \alpha},
\end{equation}
where $a_\alpha$ is the fixed scale of the theory with $\alpha$ being a theory constant; the $\alpha=1$ "simple" model is the most popular special case.
  
Reexpressing $g_s$ in terms of a rescaled Newtonian baryonic gravity $y$, we find that
\footnote{Here $\mu = 1- \left[ {g +a_\alpha\over 2 a_\alpha} +
\sqrt{\left({g -a_\alpha\over 2a_\alpha}\right)^2 + {g \over\alpha a_\alpha}} \right]^{-1}$.  The combined gravity of effective DM and baryonic matter is the actual acceleration 
${d\Phi \over dr} =g(r) = a_\alpha \left[\theta_s(y)+ \alpha^{-1} y\right]$, 
where $\theta_s(y) \equiv {2 \over 1 + \sqrt{1 + 4 y^{-1}} }$.  Note $\theta_s(y) \sim 1$ if $y \gg 1$.}
around a gas plus stellar sphere there will be an effective DM halo gravity $g_{EDM}$ or the scalar field $g_s$,
\beq\label{Phi}
 g_{EDM} \equiv g_s =  a_\alpha \theta_s(y), \qquad y \equiv { G (M_g + M_*) \alpha \over r^2 a_\alpha }
\eeq
where $M_g+M_*$ is the gas plus star mass inside radius $r$.
A remarkable result of the general class of $\mu$-function
is that there is a maximum to the scalar field gravity 
\beq g_s  \le g_{s,max} = a_\alpha. \eeq 
This means that
when the Newtonian gravity, $g_N$, is the strongest (as in the
centers of bright galaxies), the scalar field $g_s \rightarrow
a_0 /\alpha$, i.e., approaching a universal constant plateau.
This breaks down only if $\alpha=0$, corresponding to Bekenstein's toy function.
Also if $\alpha \rightarrow \infty$, the dynamics becomes purely Newtonian with a zero scalar field.  The standard $\mu(x)=x/\sqrt{1+x^2}$, which fits observations well, can be approximated by $\alpha \sim 3$ in terms of sharpness of transition from strong to weak gravity.  One can set $\alpha \sim 1-3$ to be consistent with galaxy data.
\footnote{
Based on theoretical arguments and matching observed
rotation curves, Zhao \& Famaey (2006) advocated the "simple" function with
$\alpha =1, \qquad \mu_s = {g_s \over a_0-g_s}, \qquad \mu = {g \over a_0+g}.$
This $\mu$-function is also supported by the Milky Way kinematics data,
and by the SDSS extragalactic satellite velocity distribution (Angus et al. 2007),
and is consistent with the recent rotation curve
fittings by Famaey et al. (2007a,b), Zhao \& Famaey (2006), and Sanders \& Noordermeer (2007), McGaugh (2008).  } 

The density profile of the EDM  can be derived from the Poisson equation as
\begin{equation}
 \rho_{\rm EDM}(r)={1 \over 4\pi Gr^2}{d \over dr}(r^2 g_{EDM}), \label{poisson}.
\end{equation}
Taking $g_{EDM} = a_\alpha \theta_s$ and the Newtonian gravity of the baryonic Hernquist profile (Eq.~\ref{Hernquist}), we find
\beq
\rho_{EDM} = {a_\alpha  \over 2 \pi \alpha G r} \theta_1,
\theta_1 \equiv  \left(\theta_s - {d \theta_s \over d \ln y} {r \over r+r_h}\right). 
\eeq
For small radii ($\sim$ kpc) $\rho_{EDM}$ has a roughly $1/r$ cusp because $\theta_1 \sim \theta_s \sim 1$ at centers of bright galaxies, where $g_N \sim GM/r_h^2 \sim 350 (M/1.5 \times 10^{11}\msun)^{-1.52} a_0 \gg a_0$.  In fact near the center 
$g_N \gg a_0 \sim a_\alpha$ 
even if adopting a scale length 3 times bigger than implied by the mean scaling in Chen \& Zhao (2006), and/or using a Sersic profile for the stellar distribution.   The strong gravity implies a saturated scalar field in the center: $g_s$ reaches its maximum $a_\alpha$.  
This is confirmed numerically at least for the $\alpha=1$ case (cf. Fig.~\ref{fig:daming}).  In fact near the centers of ellipticals the model predicts 
\beq
r \rho_{EDM} = {\Pi \over \alpha}, \qquad \Pi \equiv {a_0 \over 2 \pi G} \sim 130\msun\pc^{-2}.
\eeq
This is rather similiar to the case of NFW halos, but $r\rho_{EDM}$ has {\it virtually no scatter} for the MONDian scalar field in bright centers of ellipticals. 
Fig.~\ref{fig:daming} shows the scalar field is very rigid,
\footnote{The scalar field becomes half-saturated 
as soon as the overall gravity exceeds $a_0/\alpha$, where $g_s=g_N=0.5a_0/\alpha$. 
In the Milky Way this translates to a galactocentric radius of about
$(220\kms)^2 \alpha/(a_0) \sim 13 \alpha$kpc, which is the edge of the galaxy disk.} 
$g_s =(0.5-1)a_0/\alpha$, nearly incompressible on scales of 1-10 kpc, i.e., it cannot be
increased significantly by compressing the baryonic material and
increasing the Newtonian gravity.    
We can also define a maximum central pressure of the scalar field for later use
\footnote{Zhao (2007) noted
that this pressure $P_\alpha$ ultimately relates to the cosmological constant
or the vacuum energy density, which has the same order of
magnitude as $P_\alpha$}
\beq P_\alpha \equiv {a_\alpha^2 \over 4 \pi G}, \qquad a_\alpha=a_0/\alpha.  \eeq 

It is remarkable that the centers of ellipticals are all immersed in 
strong gravity, hence a universal constant scalar field, perhaps extending all the way to the central black holes of these system.
One ponders {\it the consequences of such a remarkable
uniformity and universality for central kpc regions of all bright galaxies}.  
One wonders, in particular, whether the uniformity would lead to very tight correlations, such as the $M_{BH}-\sigma$ relation.

\section{Modeling spheroid formation in a constant background gravity $a_\alpha$}

As a demonstration of the consequence of a universal scale,
we apply it to galaxy formation and scaling relations.
Unless stated otherwise we make the approximation 
\beq
g_{NFW} \sim g_{EDM} =g_s = a_\alpha \theta_s \sim a_\alpha = cst 
\eeq
This approximation means that baryons experience a rigid uniform extra field $a_\alpha$ 
on top of the Newtonian self-gravity of the stars and gas,
\beq
g(r) = a_\alpha  +  { G (M_g + M_*) \over r^2}.
\eeq   
\begin{itemize}
\item If the origin of the extra field 
is actually from NFW cusped CDM, then it should be understood that $a_\alpha$ would have 
a significant scatter between galaxies with a trend for bigger $a_\alpha$ for bigger galaxies.
\item If the origin is the scalar field in a co-variant version of MOND, then $a_\alpha$ should be viewed as a universal fundamental constant intrinsic to a theory. It is equal to $a_0/\alpha$ with zero scatter, although the value of $\alpha \sim (1-3)$ is not precisely determined at present.
\end{itemize}
Spatial non-uniformness of the background field is studied in the Appendix;
the effect can be treated crudely as a small spatial variation of $\alpha$.

\subsection{Maximum stable gas mass} 

First we construct analytical spherical models of gas and star equilibrium in MOND, and study the condition for the gas sphere to remain stable.  It is found that as we increase the core density up to a critical value, the total gas mass increases.  Once the gas core density exceeds the critical, the total gas mass  starts to decrease.

Assume a quasi-static equilibrium of a gas sphere $\rho_g(r)$ of 
sound speed $c_g(r)$ and a stellar sphere $\rho_*(r)$ of radial velocity dispersion $\sigma_*$ and 
an anisotropy parameter $\beta \equiv 1-{\sigma_{*\phi}^2 \over \sigma_*^2}$.
To model the tangential velocity dispersion and non-isothermal profile 
we introduce a velocity dispersion measure $\sigma_1$, 
\beq
\sigma_1 \equiv \xi_* \sigma_*, 
\qquad \xi_*^2 \equiv  
{d\log \rho_*\sigma_*^2 \over d\log \rho_*} -{d\log r \over d\log \rho_*}\beta.
\eeq

Define $\xi$ to be the ratio of thermal pressure $\rho_g\sigma_g^2$ to the 
opacity-induced pressure $(\rho_g \sigma_g^2)/\xi$ on the dusty gas
sphere (not the stellar sphere), countering the gravity.
The velocity dispersion $\sigma_g(r)$ is related to the sound speed by  
$c_g^2=\sigma_g^2 {d \log \rho_g \sigma_g^2 \over d\log \rho_g}$.
In equilibrium the gas-stars mixture satisfies the equations \beqa
g(r) &=& {  c_g^2 \over r} \left[{-d\ln[\rho_g(r)] \over d\ln r}\right] (1+\xi^{-1})\\\nonumber
&=& {\sigma_1^2 \over r}\left[-{d\ln[\rho_*(r)] \over d\ln r}\right],\\
4\pi r^2 &=& {dM_g(r) \over \rho_g(r) dr} = {dM_*(r) \over \rho_*(r) dr}. \eeqa

The conversion of gas into stars also needs to be modeled to be realistic.
The simplest solution of the above eqs. would be a model
where {\it stars trace the gas radial distribution}, so we have
\beq
{\rho_*(r) \over \rho_g(r)}={M_*(r) \over M_g(r)} = {f_*(t) \over 1-f_*(t)},
\eeq
where the position-independent factor $f_*(t)$ is the fraction of gas
formed into stars at time $t$.  Such a solution is possible if the
feedback ratio $\xi^{-1}$ is regulated by star formation, and related to the temperature of the gas by
\beq
(1+\xi^{-1})c_g^2 = \xi_*^2 \sigma_*^2 = \sigma_1^2 =cst.
\eeq
Note that the feedback parameter $\xi$ and the gas sound speed $c_g$ are not required to be rigorously independent of radius, although certain combination of these two is.  We do not require the stellar component to be exactly isothermal or isotropic either.  

We compute the gas equilibrium for a range of core pressures $p(0)$
(see Figure~\ref{fig:gasden}) after 
rewriting the equations in term of the dimensionless mass $m$, radius $u(m)$ and rescaled density $p(m)$ (see Appendix).  
We find that the gas density generally
falls off monotonically with radius or mass.  All models have finite
mass out to infinite radius.  The density drops steeply with radius
due to the deep linear potential well of the background gravity,
hence the mass converges quickly.
Curiously there is also a maximum $m_{max} \approx 4.3$
in the total rescaled mass as $p(0)$ increases.
This happens at a critical core density or pressure
\beq p(0) = {\rho_0 \sigma_1^2 \over P_\alpha} \approx 30
\eeq
above which the gas density $\rho_g$ of a parcel of gas $dM_g$
no longer increases monotonically with an increase of the central pressure,
and in fact the total mass will decrease with increasing
$p(0)$ after it reaches the maximum value.

These limits on gas central pressure and total mass are examples of
the instability first discussed by Elmegreen (1999).
A gas sphere above a certain critical mass
\beq
M_{max}  \approx 4.3  \left({\sigma_1^4 \over a_\alpha G}\right),
\eeq
or above a critical central gas density or pressure,
does not have stable solutions; adding a tiny amount of gas would lead to collapse.  It is interesting to speculate that bulge
formation originates from such a gas instability.

For a sphere of gas plus stars at the critical mass, we 
integrate the density numerically, and find the central surface mass density 
\beq\label{I0}
S(0) \sim 2 a_\alpha/G \sim 1500 \alpha^{-1} \msun\pc^{-2},\eeq 
insensitive to the initial gas
velocity dispersion $\sigma_g$ and the feedback ratio $\xi^{-1}$.
We also fit our numerical solution of the gas mass profile by a Sersic-like distribution.
The gas mass profile is related to the dimensionless volume density profile $j(u)$ via
\beq 
{M_g(r,t) \over (1-f_*(t)) M_{max}} =
 \int_0^u j(u) (4 \pi u^2 du), ~ u \equiv {r
\over \sigma_1^2 a_\alpha^{-1}}.
\eeq
Prugniel \& Simien (1997, eq B6) suggested an approximated (valid to 5\% accuracy typically) form $r^{-1+0.6/n}\exp(-r^{1/n})$ for the deprojected volume density of a Sersic profile.   
We fit $j(u)$ numerically to such a S{\'e}rsic profile of volume density, and find a $n=1.2$ profile
works well, i.e.,
\beq
\qquad j(u) \approx 0.14 u^{-1/2}
\exp(-1.6 u^{1/1.2}). \eeq
The normalisation is such that $\int_0^\infty 4\pi j(u) u^2 du=1$.

\subsection{Properties of the stellar component formed}

In the simplest scenario gas might turn adiabatically into
stars while maintaining the density profile
at the critical density and mass.
Eventually $f_*=1$ when the gas is exhausted by star formation,
we have a stellar system with a nearly Sersic profile of 
high central surface brightness.  The density slope
is shallower/steepr than isothermal inside/outside the radius $r \sim 0.3 \sigma_1^2 a_\alpha^{-1}$.

Our model also resembles real spheroids since they satisfy 
the Faber-Jackson-like relation $M_*^\infty \sim
\sigma_*^4$ between the total mass and stellar velocity dispersion (Faber
\& Jackson 1976),  
\beq\label{FJ}
 M_*^\infty = {4.3 \sigma_1^4
\over a_\alpha G } = 4.3 \xi_*^4 \alpha \times 10^{11} \Msun \left({\sigma_* \over
200\kms}\right)^4. \eeq 
Considering small deviations from  isothermal and isotropic $\beta \sim 0$
stellar distribution, $\sigma_* = \sigma_1/\xi_*$ can be treated as effectively certain mean projected stellar dispersion.  
Note that our result differs in detail
from the MOND virial theorem
$M_* = {81 \over 4G a_0}\sigma_*^4$ (Sanders \& McGaugh 2002)
which applies to an isotropic isothermal stellar system in deep-MOND (see Appendix).
Our bulges and ellipticals are clearly in a mild or high acceleration regime.

\subsection{Correlations of Black Hole, Stellar Spheroid and Starburst}

Similar to the introduction, we assume 
the momentum deposit rate of photons, ${2 \over c}( L_{SF} + L_{BH} )$ 
drives a force acting on the gas.
Assume the BH grows exponentially by Eddington accretion from a seed before 
reaching the balance $L_{BH} = L_{SF}$.  After this point, 
the BH mass stops growing and we set $L_{BH}=0$, and 
the gas mass and the $L_{SF}$ will then exhaust exponentially as the starburst continues.  
Assuming at the turning point, the BH and SF each provide half of the total feedback force, which is a fraction $(1+\xi)^{-1} \sim 1$ 
of the radially-pointing gravity $g(r)$, we obtain 
\beq {2 \over c} L_{BH} = {2 \over c} L_{SF}  
 = {\bar{g} \over 2 (1+\xi)} M_{g} 
\eeq
where $M_g =(1-f_*) M_{max}$ is the total gas mass, and 
the mass-averaged gravity 
\beq
\bar{g} \equiv M_g^{-1} \int g dM_g  \sim 2a_\alpha
\eeq
is computed by an integration of mass for the density model with the critical mass.
This result is robust for a range of the polytropic index of the gas (see Appendix). 

Observations reveal that some gas rich local spirals have $f_* \sim 0.5 \pm 0.2$.  
The stellar fraction is
$f_* \sim 1$ for local bulges and elliptical galaxies.  This is much higher than
in their $z \sim 3$ progenitor starburst galaxies $f_* \sim 0.02$ 
(Alexander et al. 2005; Genzel et al. 2006; Kennicutt 1998, Solomon \& Vanden Bout 2005). 
We shall adopt $M_g =(1-f_*) M_{max}=(0.5-1)M_*^\infty$, 
and $\bar{g} \sim 2 a_\alpha \sim (1-2) \times 10^{-10}{\rm m s}^{-2}$,  
and $(1+\xi)^{-1} \sim 1$ as long as the feedback $1/\xi$ is not too small (Murray, Quartet \& Thompson 2005).  We obtain in the end
\beq 
{M_{BH} \over 2 \times 10^8\Msun} 
= { L_{SF} \over 6 \times 10^{12}\Lsun } 
= {M_*^\infty \xi' \over \alpha \times 10^{11}\Msun} 
= \left({\xi_*' \sigma_* \over 200~\kms}\right)^4 , \eeq 
where $\xi'=(1-f_*)/(1+\xi) \sim 0.5-1$, consistent with the observed Magorrian relation 
$M_{BH} \sim 0.005 M_{\rm bulge}$, and $\xi_*' = (4.3 \xi')^{1/4} \xi_* \sim 1$.
The scatter in $\xi_*'^4$ is about factor of a few each way, 
consistent with the narrow scatter seen in the observed relation of 
$M_{BH} \sim (1-4)\times 10^8\msun ({\sigma_*/200\kms})^4$ (Ferrarese \&
Merritt 2000; Gebhardt et al. 2000; Tremaine et al. 2002).

\subsection{Hydro simulations of gas collapse in a fixed background potential}

To confirm the analytical results on the gas instability, we investigate the threshold of gas collapse numerically. We consider the gravitational collapse of gas within a rigid dark matter halo using a hydrodynamic simulation.  The gas which forms the bulge is embedded in a
uniform external field from the dark matter potential.  The simulation represents a simple treatment of the CDM halo gravity or the MOND scalar field.

This simulation was performed using the Lagrangian fluids code,
Smoothed Particle Hydrodynamics (SPH). The code is described in Bate,
Bonnell \& Price (1995), and is based on the version by Benz et. al. (1990). The simulation uses $10^5
$ particles, which are initially distributed randomly within a sphere
of radius 15 kpc. The gas is initially at rest, with a temperature
of $10^6$ K. The mean molecular weight is $1.2$, thus the sound
speed of the gas is 83 km s$^{-1}$. The dark matter halo is included
by way of a fixed linear external potential. The potential per unit mass is $-a_\alpha  \mathbf{r}$ such that there is a universal acceleration of $a_\alpha=10^{-10}$ m sec$^{-2}$ towards the centre of the galaxy. The gas is
subject to both the external potential and self gravity during the
calculations.

With these parameters, the critical mass is $1.54 \times 10^{10}$~M$_
{\odot}$. We first set the total mass of the sphere to 0.0065 M$_
{max}$ ($10^8$~M$_{\odot}$), so the gas should be stable against
gravitational collapse. We then evolved the gas for 12 crossing times,
during which time the gas settles into equilibrium according to the
dark matter potential. The mass was then doubled (by doubling the
mass of each particle) and the calculation resumed. By 5 crossing
times, the gas had again settled into equilibrium. This process was
repeated, doubling the mass each time the gas reaches equilibrium,
until runaway gravitational collapse occurs.  Fig.~\ref{fig:snap}
shows snapshots of the radial profile of the gas pressure 
just below (upper panles) and just above (lower panels) the critical mass.  
A collapse of gas is evident from the density cusp in the final state.  

In Fig~\ref{fig:hydro},
a 1 D projection of the column density (along the $x$ axis)
is plotted versus radius. During the first run, the distribution of
gas changes from a uniform profile, to a profile corresponding to the
dark matter halo. Over the subsequent runs, there is little change in
the distribution whilst the total mass remains less than the critical
mass. However when the mass exceeds the critical mass, the cloud is
gravitationally unstable. Gravitational collapse occurs and the
density at the centre of the cloud continuously increases. The
calculation stops, the final profile shown as the thickest line in
Fig.~\ref{fig:hydro}; the slightly subcritical gas can be approximated by 
a Sersic law with $n =1$ (i.e., exponential). 

Galaxy formation clearly involves more than just spherical collapse.  The rapid spherical collapse phase might produce the bulge and the black hole.   This phase is likely followed by a gradual phase of episodes of minor mergers, the dense galaxy will likely acquire a diffuse stellar halo of higher angular momentum material.   One might expect a shallower outer profile than $n=1$ in the end depending on the amount of stars accreted.  Indeed real galaxies have a range in Sersic index, which is correlated with the stellar mass of the galaxy (Caon et al. 1993, Desroches et al. 2007).  Bulges and pesudo-bulges typically have profiles with $4 \ge n \ge 1$, and the most massive elliptical galaxies have profiles with $n \ge 4$.  Some discrepancy with observations is expected  since a constant
scalar field is a poor assumption at large radii.  


\section{Summary}

In short, combining the results of analytical models and numerical simulations, we argue that gas collapse and BH feedback inside the effective DM halos of MOND can produce galaxies with realistic scaling relations.  Better than in Newtonian NFW halos, 
the Faber-Jackson and BH mass-velocity dispersion scaling relations
are recovered with narrow scatter thanks to the fact that there is a 
redshift-insensitive and luminosity-insensitive universal scale
of gravity in the high-z gas-rich starburst galaxies and present day ellipticals. 
Many feedback processes (e.g., Silk \& Rees 1998, King \& Pounds 2003, Cen 2007, Xu, Wu, Zhao 2007) invoked in the CDM context could be important  in MOND context as well, and   
could change predictions of the state of the art MOND N-body simulation (Nipoti et al. 2008) and hydro simulations (Tiret \& Combes 2008).

In the context of a simplified picture of TeVeS, a nearly-isothermal gas sphere can collapse and trigger a starburst if the gas central pressure is above a universal threshold. This condition is likely synchronised throughout the universe, consistent with the observed epoch of starbursts.  We also recover the $M_{BH}-\sigma_*$ relation, if the gas collapse is regulated or resisted by the feedback from radiation from the central BH.

While the pristine CDM halos give acceleration of typically the order $(1-3)a_0$,
there are two important differences with the universal scale in MOND.  In CDM, this scale would exhibit a large variation due to scatter in halo concentration (Milgrom 2002).  Secondly, as the galaxy bulge forms, the central part becomes baryon-dominated in Newtonian gravity and the CDM is adiabatically compressed to higher densities without a theoretical upper limit unless DM is made of neutrino or its sterile partners (cf. Angus 2008, Zhao 2008).  The collapse threshold could rise as the background gravity $a_\alpha$ increases.   In contrast, in the TeVeS picture the scalar field $g_s$ within all bright galaxies stays close to $a_\alpha$ throughout galaxy formation. This value $a_\alpha$ is a universal constant in MOND.  Hence we expect a universal theshold and synchronised formation of galaxy spheroids at high redshift, perhaps producing star burst galaxies in TeVeS.  The mechanism would work less well in the CDM paradigm.

Speculating beyond our models, 
elliptical galaxies are often triaxial in their stellar distribution and potential.  If collapse happened above the threshold described here,
the uneven distribution of angular momentum would lead to a triaxial equilibrium potential.  This expectation is consistent with 
N-body simulations of collapse in CDM halos (Navarro et al. 1996) and in MOND (Nipoti et al. 2007).  Indeed, elliptical galaxies with a mild $r^{-1}$ cusp in stellar light are allowed to exist in self-consistent triaxial configuration in CDM (Capuzzo-Dolcetta et al. 2007) and in MOND (Wang et al. 2008).  

Corrections are expected for low surface brightness galaxies since the scalar field in these systems are far from being saturated. So $\bar{g} \sim w a_\alpha$, where $w \le 0.3$, and we expect the faint dwarf spheroidal mass and their central BH mass to reduce by a factor $w$ and $w^2$ respectively compared to a blind application of Faber-Jackson and $M_{BH}-\sigma$ relations.  No correction is expected for M32-like compact dwarf ellipticals, which are observed to have bright stellar nuclei and BHs.   Our prediction that  the BH mass per unit stellar luminosity $M_{BH}/L_*$ is lower in dwarf spheroidals than dwarf ellipticals could be  checked by sensitive searches for (central tracers of) massive BHs in dwarf spheroidals, which is challenging observationally.

Environmental effects could lead to corrections to the Faber-Jackson relation.  Member galaxies in the center of a rich galaxy cluster have orbital accelerations of $\sim 2 a_0$, which makes it easier to saturate their MOND scalar field due to the external field effect; any gas rich system would be more vulnerable to radial collapse into a high surface brightness triaxial galaxy in the center of a galaxy cluster than in the surrounding or in the field (Wu et al. 2007, 2008).  The external field effect 
is similar to making $a_\alpha$ smaller.  Interestingly, eq.~(\ref{I0}) and~(\ref{FJ}) can be combined to form 
\beq
{M_*^2 \over L_*^2} \sim {\sigma_*^4 \over I_* L_*}{ 9\xi_*^4 \over G^2} 
\sim  
\left({\sigma_* \over 200\kms}\right)^4 \left({1500L_\sun\pc^{-2} \over I_*}\right){4.3\xi_*^4 10^{11} L_\sun \over L_*},
\eeq
where $I_* = S_*/(M/L)$ is the surface brightness of stellar distribution of surface density $S_*$. 
This relation is very similar to the fundamental plane relation of ellipticals (e.g. Binney \& Merrifield 1998)
\beq
L_* \propto \left({\sigma_*^4 \over I_*}\right)^{0.7}, 
\eeq
if we accept that $M_*/L_* \sim L_*^{0.2}$.  Equivalently 
we can write the galaxy size $r_e \propto S_*^{-1} \sigma^2$, which is 
very close to the lensing mass fundamental plane $S_* \propto \sigma^2/r_e$
determined by GR-based strong lensing results, 
$S_*^{0.93} \propto \sigma^{1.93}/r_e$ (Bolton et al. 2007);
the MONDian corrections to the lensing mass within the Einstein radius are generally mild
(Zhao, Bacon, Taylor, Horne 2006, Shan, Feix, Famaey, Zhao 2008).  On the other hand, 
environmental effects would not correct the BH-velocity dispersion relation because
\beq
M_{BH} \propto \sigma_*^4, 
\eeq
which is independent of the parameter $a_\alpha$.

The BH-bulge relation may have been frozen at high redshift; 
some scatter might be 
caused by gas rich mergers, and minor feeding of the BH through a rotating gas-rich bar later on, which might evolve secularly into a bulge (Tiret \& Combes 2007).  Disk galaxies could have built up their disks from high-angular momentum gas well after the bright bulges form by radial collapse.  The formation of the disk part will not fuel the central BH significantly.  This may explain why the BH mass is tightly related to the bulge and its velocity dispersion rather than with the total baryonic mass and the terminal circular velocity in disk galaxies; the latter two are correlated themselves by the Tully-Fisher relation, which is built in the Lagrangian of covariant MOND because  virtually no scatter from this relation is observed in field galaxies (cf. Wu et al. 2007).

\acknowledgments

We acknowledge the support of NSFC fund (No.10473001 and
No.10525313) and the RFDP Grant (No.20050001026) to Xuebing Wu and BXX,
and partial support from NSFC Grant to HSZ (No. 10233040) and PPARC.  HSZ thanks Da-ming Chen and Xufen Wu especially for technical assistance, and Xue-Bing Wu, Benoit Famaey, Gianfranco Gentile and Ralf Klessen, Keith Horne, Simon Driver, Ian Bonnell for discussions.  CLD's work is conducted as part of the award `The formation of stars
and planets: Radiation hydrodynamical and magnetohydrodynamical
simulations', made under the European Heads of Research Councils and
European Science Foundation EURYI (European Young Investigator)
Awards scheme, and supported by funds from the Participating
Organisations of EURYI and the EC Sixth Framework Programme.

\begin{figure}
\includegraphics[height=16cm, width=16cm,angle=0]{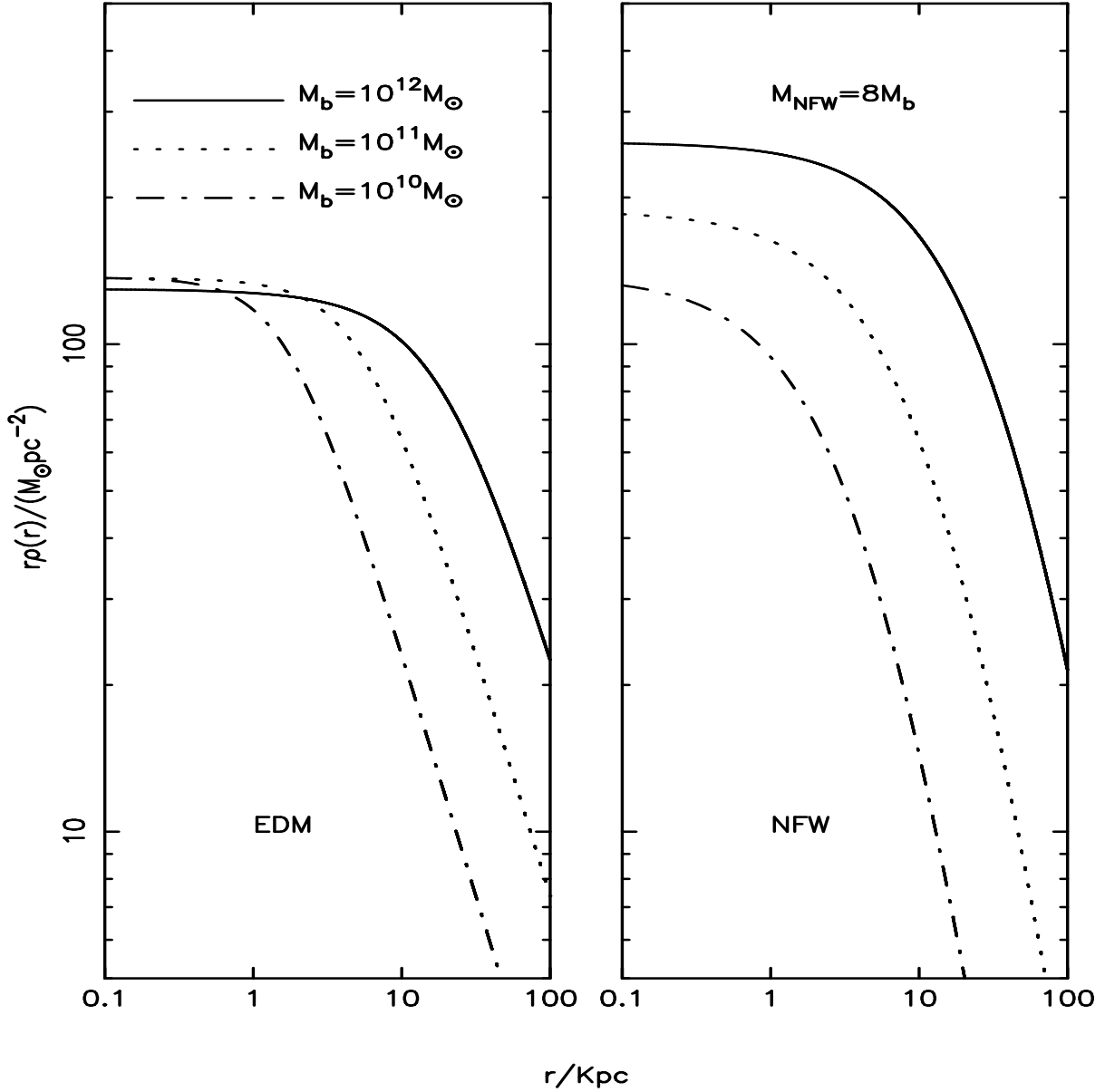}
\caption{The mass distribution of Effective DM halos (left
panel and assuming an $\alpha=1$ MOND) and NFW Cold DM halos (right panel) in terms of $r\rho(r)$.  Note the nearly universal central density, especially in MOND.  Not shown here is a further 
factor of 10 scatter intrinsic to NFW CDM halo density, and 
another factor of $\sim 10-100$ upward correction if elliptical galaxies form adiabatically and compress CDM.  For MONDian $r \rho_{EDM}$, four lines represent
$M_b=(10^{12}, 10^{11}, 10^{10})M_{\sun}$, respectively, as indicated. For
CDM $r \rho_{\rm NFW}$, we take the corresponding value of the halo mass as $M_{\rm NFW}=8M_b$.   }\label{fig:daming}
\end{figure}

\begin{figure}
\includegraphics[height=16cm, width=16cm,angle=0]{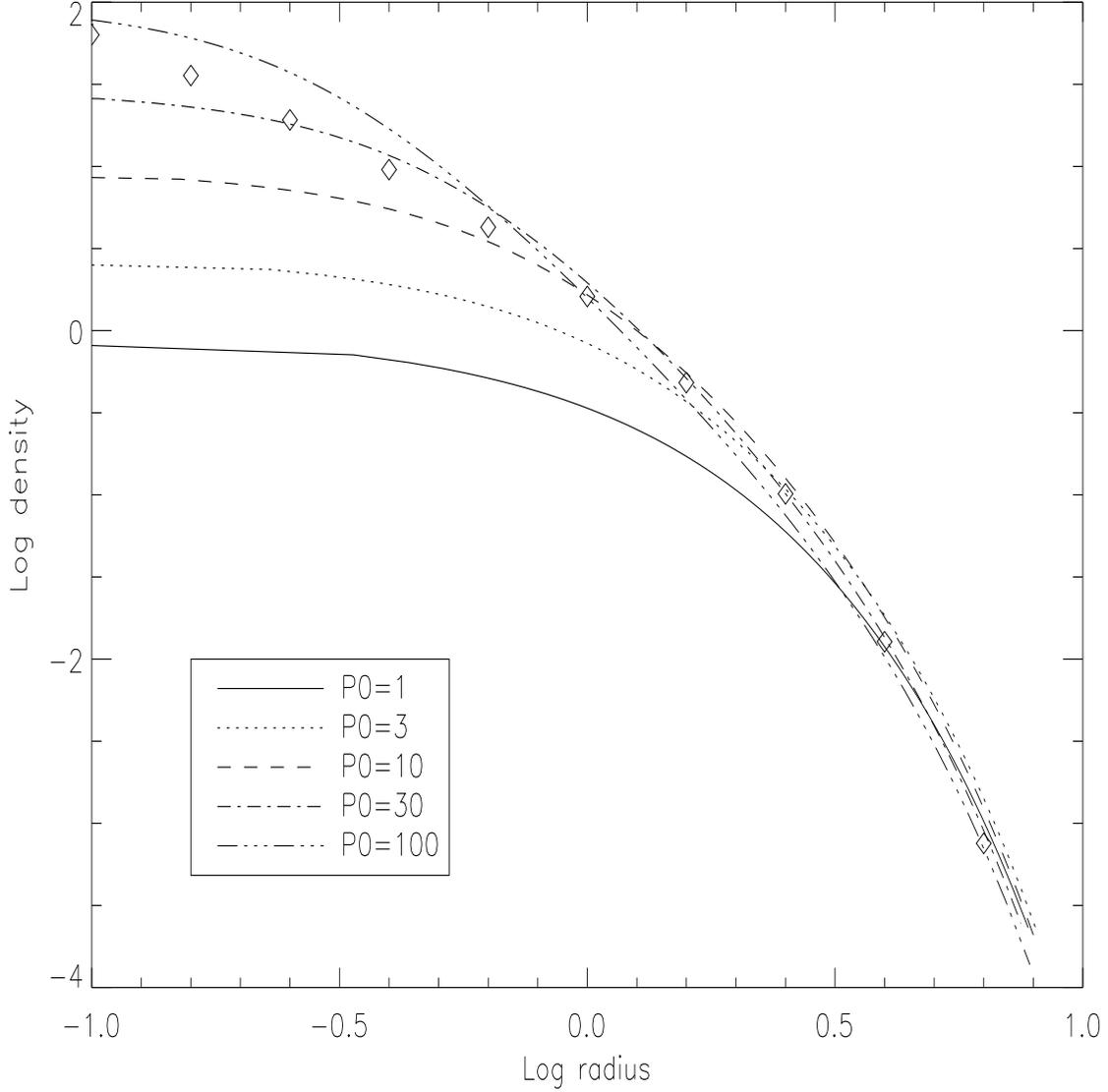}
\caption{Mass distribution of
a gas sphere embedded in a rigid background uniform gravity $a_\alpha$ for models (from bottom to top at small radii)
with increasing dimensionless central gas pressure
$p_0=1,3,10$ (bottom three), and $p_0=30,100$ (top two).
The axes are $\log p(m)$ vs the logarithm of the rescaled radius
$u(m) = {r \over \sigma^2/a_\alpha}$,
where $m = G a_{\alpha} \sigma_1^{-4} M_g$ is the rescaled baryonic mass, 
$p(m) = {\rho_g(M) \sigma_1^2 \over a_\alpha^2/(4 \pi G)}$ is
the rescaled gas density or pressure.
Note how the high-$p_0$ lines are above/below the lower $p_0$ lines at small/large radii respectively.  This {\it reversal} is a feature of reaching a maximum in
total gas mass at the critical pressure ($p_0 \sim 30$).
Also shown is a S{\'e}rsic $(n=1.2)$ profile (diamonds).
}
\label{fig:gasden}
\end{figure}

\begin{figure}
\includegraphics[height=16cm, width=16cm,angle=-90]{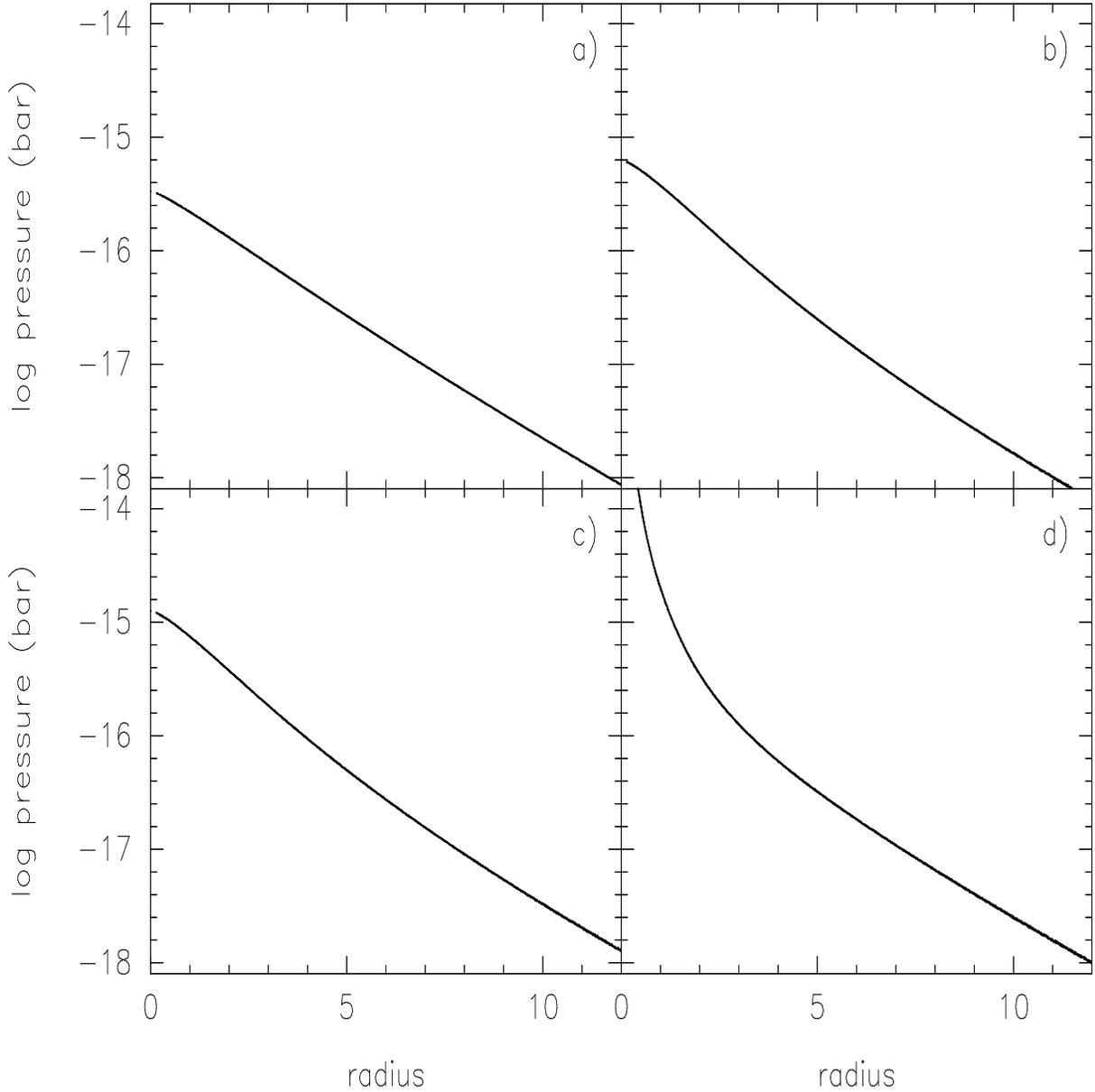}
\caption{shows the gas pressure radial profile of the 3D hydro simulations in a fixed background field $a_\alpha=a_0$.  
Panels (a) and (b) are the "before" and "after" evolution of a gas sphere of low mass $M_g=3.57$ in 
units of $\sigma_1^4/(Ga_\alpha)$ (see text for physical units), which is below the critical value $4.3$.  
Panel (c) and (d) show the prominent development of 
a gas density cusp for a high mass $M_g=7.14$, exceeding the critical gas mass.
Note the gas profiles are nearly exponential 
and are consistent with a Sersic index $n$ about 1-1.2 before it becomes unstable.
In d), the pressure in the centre exceeds the 
critical pressure, which is $\sim 3.6 \times 10^{-15}$ bar.}\label{fig:snap}
\end{figure}

\begin{figure}
\includegraphics[height=16cm, width=19cm,angle=0]{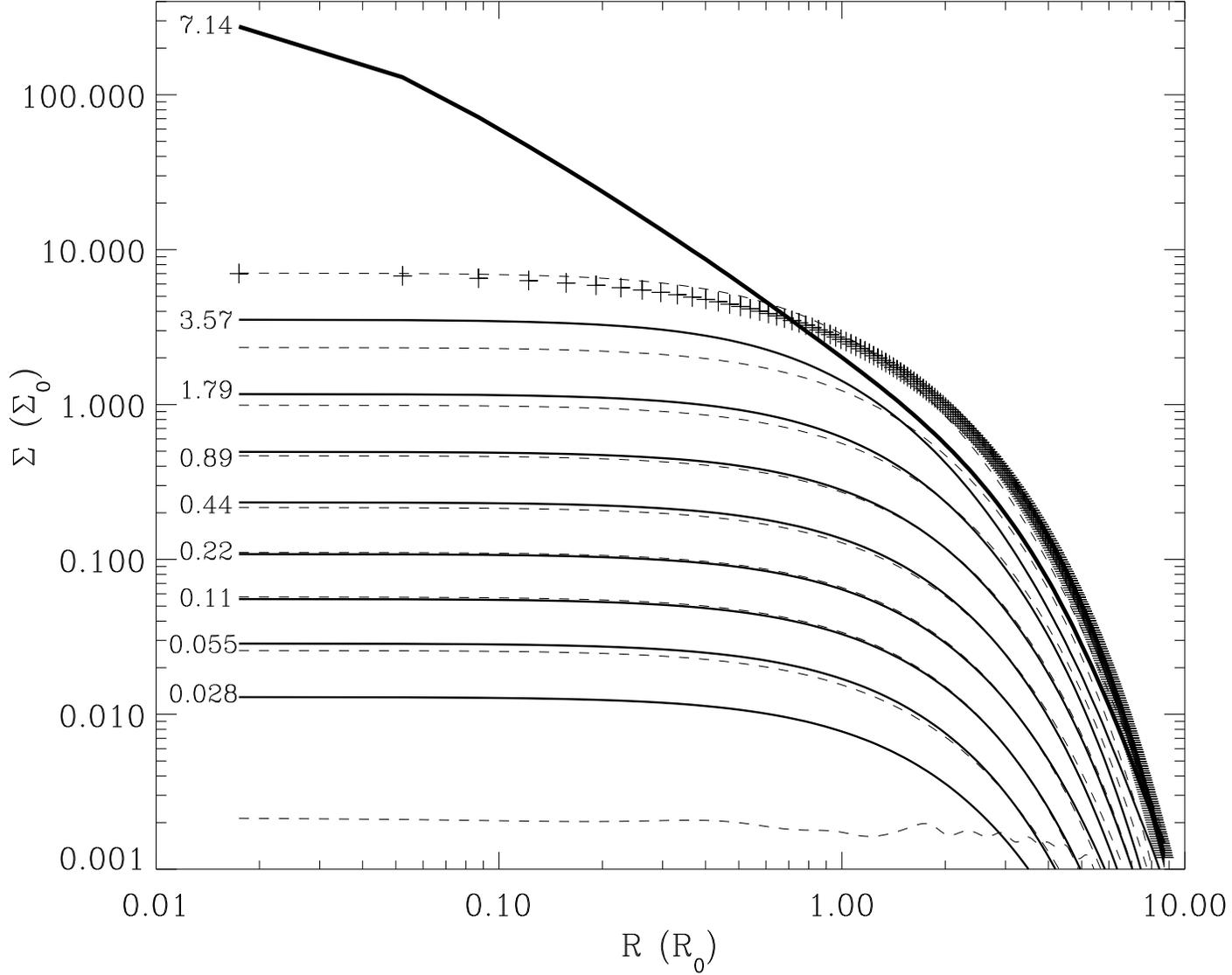}
\caption{Shows hydro simulation in a fixed background field $a_\alpha=a_0$.  The
numbers on the left are total gas mass $M_{gas}$ in units of $\sigma_1^4/(G a_\alpha)$.
The dashed lines represent the input gas profile 'before' the
evolution, and solid 'after' the hydro evolution. The timescale for
each evolution is 5 crossing times, except for the first run from a
uniform distribution, which was evolved for 12 crossing times.
The radius is in unit of $R_0 \equiv \sigma_1^2/a_\alpha$, and the projected density in units of $\Sigma_0 \equiv a_\alpha/(2\pi G)$.  Instability happens when the critical mass exceeds 4.3 in these units, or the surface density exceeds about 10 in these units.  
The slightly subcritical gas can be approximated by 
a Sersic surface density with $n =1$ (plus signs). }\label{fig:hydro}
\end{figure}

\begin{figure}
\includegraphics[height=16cm, width=16cm,angle=0]{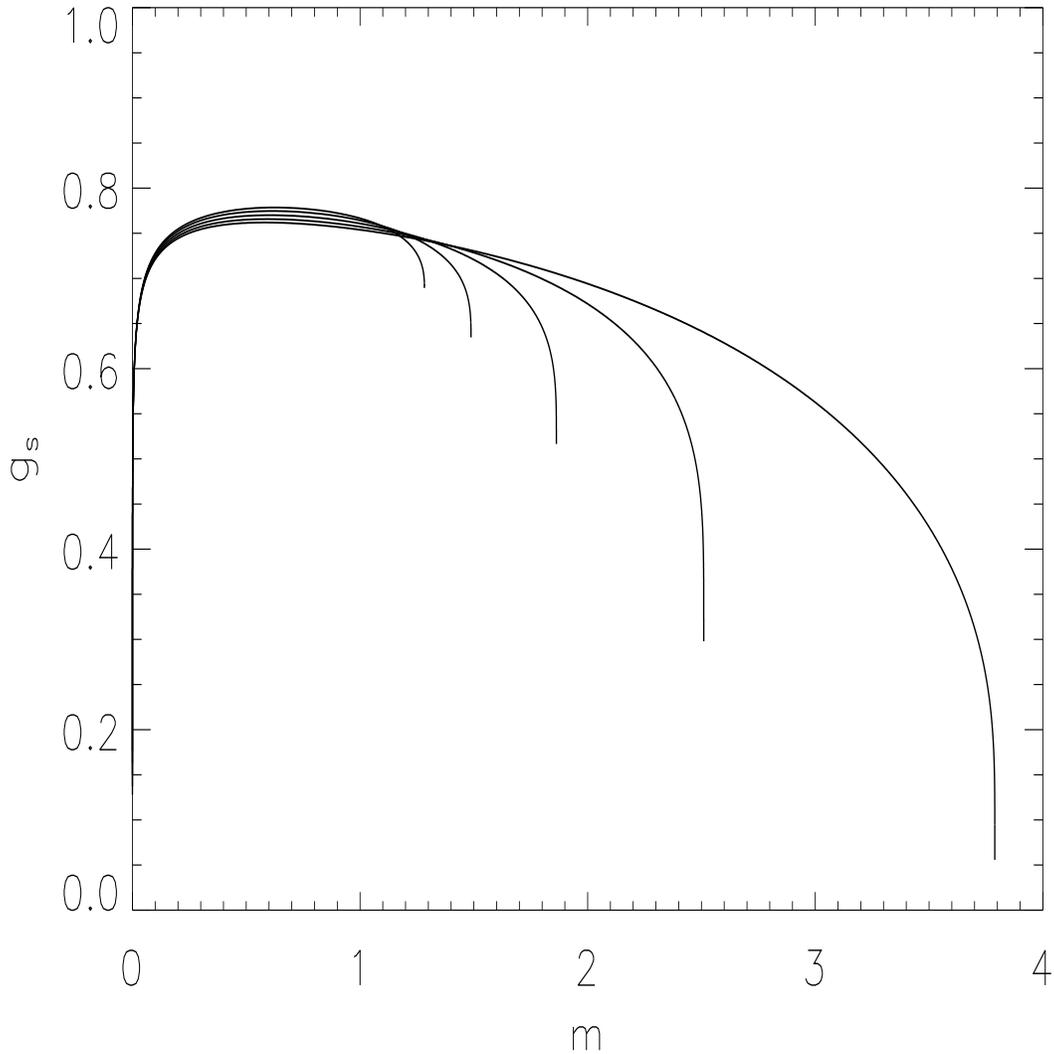}
\vskip 3cm
\caption{Effective dark matter (scalar field) gravity $g_s$ in units of MOND $a_\alpha$ ($\alpha=1$) as a function of enclosed gas mass in units of $\sigma_g(0)^4 /(Ga_\alpha)$ for a non-isothermal polytropic gas sphere of central velocity dispersion $\sigma(0)$ and with $\gamma=1.1$ (rightmost curve), 1.2, 1.33, 1.5, 1.66 (leftmost curve).  All the curves have central pressure $~30 P_\alpha$, and the pressure drops to zero at the last points of the curves (truncation radii of the polytrope).}\label{fig:poly}
\end{figure}

\appendix

\section{Mean gravity and scalar field in MONDian polytropic gas models with/without feedback}

\subsection{Constant background gravity model with Feedback}
We rewrite the hydrostatic equations using the dimensionless quantities
 \beq
 m \equiv {M_g+M_* \over \sigma_1^4 a_\alpha^{-1} G^{-1}},~
 u \equiv {r \over \sigma_1^2 a_\alpha^{-1}},~
 p \equiv {\rho_g(r) + \rho_*(r) \over P_\alpha \sigma_1^{-2} }
\eeq
and expressing the rescaled mass $m$ as the independent coordinate,
the problem is recast to solving the pair of dimensionless ODEs
\beq\label{ode}
-{u^2 dp(m) \over dm} = \theta_s + {m \over u(m)^2}, \qquad
 {u^2 du(m) \over dm} = {1 \over p(m)} .
\eeq
For the constant background gravity models, we set $\theta_s=1$.
For each value of $p(0)$, the density profile under the
hydrostatic equilibrium can be completely determined with the
initial conditions at the center for the radius $u(0)=0$ and the rescaled density $p(0)$.  Results are shown in Fig.~\ref{fig:gasden}. 

\subsection{Polytropic self-gravitating models}

The above simplified models
apply only if we can treat the scalar field as constant, and the velocity dispersion as isothermal.  These are not rigorous for real gas.  To check the robustness of our results we have also computed numerically more realistic models where we solve for a self-gravitating polytrope $\sigma_g^2 \rho_g^{1-\gamma} =cst$ in rigorous self-consistent equilibrium in MOND gravity.  There are no SF and BH in these models.  
Nevertheless the pressure force on the gas sphere
\begin{eqnarray}
F &=& \int_0^{\infty} (\rho_g \sigma_g^2) d (4\pi r^2) \\
  &=& \int -(4\pi r^2) d (\rho_g \sigma_g^2) \\
  &=& \int (4\pi r^2) \rho_g g dr \\
  &=& \int g(r) dM_g 
\end{eqnarray}
where we applied integration by parts, 
\footnote{Alternative expressions are  
$F = \bar{g} M_g = \int {V_{cir}^2 \over r} dM_g = \int {2 \sigma_g(r)^2 \over r} dM_g$, which applies to non-isothermal gas as well.}
and the hydrostatic equilibrium equation
\beq
-{d(\rho_g\sigma_g^2) \over dr} = g \rho_g.
\eeq 

The radial distribution is modelled in the style of Sanders (2000), but here 
with the simple scalar field interpolation function 
$\mu_s(g_s)=g_s/(a_0 - \alpha g_s)$, so 
\beq
g = a_\alpha \left[ {m \over u(m)^2} +  \theta_s({\alpha m \over u(m)^2}) \right], 
\eeq
where $m$ and $u(m)$ are the dimensionless mass and radius, 
\beq
m \equiv { G M(r) a_\alpha \over \sigma_1^4},  
\qquad u(m) \equiv  {r a_\alpha \over \sigma_1^2},
\eeq
where $\sigma_1$ is a characteristic dispersion of the polytrope.

For models with $\alpha=1$, the 
total gravity $g = (1+\mu_s) g_s = {a_0 g_s \over a_0 -g_s}$, 
we computed (see Fig. \ref{fig:poly}) the scalar field strength for models with a central pressure $\sim 30P_\alpha$.  These models have a finite radial truncation, where the polytrope temperature $\sigma_g^2$ falls to zero.  

The profiles vary with the polytropic index $\gamma$, a value typically between 1 and 5/3 for real gas.  Nevertheless the scalar field is roughly constant even in these more realistic models where the velocity dispersion is not isothermal.  Numerical integration finds that the mass-averaged gravity 
\beq
\bar{g} \equiv  {1 \over M_g} \int g(r)  dM_g  \sim 2 a_\alpha,
\eeq 
where we used the approximation that at half-mass radii, $g_s \sim 0.66a_\alpha$, $g_N = \mu_s g_s \sim 2g_s$, and $\mu={\mu_s \over 1+\mu_s} \sim 0.66$.  So our simplistic treatment should be a good guide to real galaxies in MOND;
extending our self-consistent hydrodynamics code for Fig.~\ref{fig:snap} to incorporate a live MOND scalar field would allow us to check the universality of the threshold here. 

\subsection{Consistency with earlier models}

As a consistency check with earlier MOND models, we note that 
our finding of a maximum scalar field acceleration 
\beq
|g - g_N| \le a_\alpha ={a_0 \over \alpha}
\eeq
is consistent with Brada and Milgrom (1994)'s first finding of a maximum effective halo acceleration in many MOND $\mu$-functions, especially the standard $\mu$ function.  Fig.~\ref{fig:daming} argues the scalar field is largely a constant in the bulge regions (1-10 kpc).
We noted some similarity of a NFW halo to the MONDian scalar field with the $\mu$-function of Angus et al. (2006), especially if $\alpha=1$.  The scalar field in the TeVeS picture 
can be simply replaced by $g_s \rightarrow |\nabla {g_{00} c^2 \over 2}| - {G M \over r^2}$ in other co-variant metric versions of MOND (e.g., Zhao 2007); the counterpart in the modified inertia picture is less clear.     
Earlier authors have also noted the similarity of the standard $\mu$-function with
Pseudoisothermal halos (Sanders and Milgrom 2005), where the halo acceleration goes up to a maximum as a solid body, and then drops.
Several literatures claim that NFW halos produce somewhat higher than $a_0$ accelerations
(e.g. Fig.8 of McGaugh 2004), but the comparison of NFW and MOND has been as systematic as done here.   Works of Famaey and Binney (2004) and Sanders and Noordmeer (2007) show that real data tolerates the $\alpha=1$ $\mu$-function. 
Using the standard $\mu$, Milgrom (1984) and Sanders (2000) also note a maximum mass of an isothermal self-consistent distribution in MOND, about $(15-20) {\sigma_1^4 \over a_0 G} \sim 4.3 \alpha {\sigma_1^4 \over a_0 G}$ if we approximate $\alpha \sim 3$.  Their assumption is different because their scalar field is not fixed as a constant, but has a peak value about $a_0/\alpha \sim a_0/3$.   While differing in details, these earlier analysis are qualitatively consistent with our finding of an maximum scalar field of order $a_0/\alpha$ in MOND.   

\end{document}